\documentclass[epj]{svjour}
\usepackage{mwe}
\usepackage{widetext}
\usepackage{graphics}
\usepackage{color}
\usepackage{graphicx}
\usepackage{dcolumn}
\usepackage{bm}
\usepackage[utf8]{inputenc}
\usepackage{gensymb}
\usepackage{latexsym}
\usepackage{latexsym}
\usepackage{amssymb}
\usepackage{amsmath}
\usepackage{amsfonts}
\usepackage{nicefrac}
\usepackage{float}
\usepackage{booktabs}
\usepackage{siunitx}
\usepackage{slashed}
\usepackage{booktabs}
\usepackage{siunitx}
\usepackage[table]{xcolor}
\usepackage[mathscr,scaled=1.15]{urwchancal}
\DeclareFontFamily{OT1}{pzc}{}
\DeclareFontShape{OT1}{pzc}{m}{it}%
{<-> s * [1.15] pzcmi7t}{}
\DeclareMathAlphabet{\mathpzc}{OT1}{pzc}{m}{it}
\newcommand{\flaligne}[1]{\begin{flalign} #1 \end{flalign}}
\newcommand{\bracket}[1]{\left( #1 \right)}
\begin{document}
\title{Radially excited pion: electromagnetic form factor and the box contribution to the muon's $g-2$}
\author{Angel S. Miramontes\inst{1,2}, K. Raya\inst{3}, A. Bashir\inst{2,3}, P. Roig \inst{1,4} \and G. Paredes-Torres \inst{2}
}                     
%
\institute{Department of Theoretical Physics and IFIC, University of Valencia and CSIC, E-46100, Valencia, Spain \and Instituto de F\'isica y Matem\'aticas, Universidad Michoacana de San Nicol\'as de Hidalgo, Morelia, Michoac\'an 58040, Mexico  \and Department of Integrated Sciences and Center for Advanced Studies in Physics, Mathematics and Computation, University of Huelva, E-21071 Huelva, Spain \and Departamento de F\'isica, Centro de Investigaci\'on y de Estudios Avanzados del IPN, Apdo. Postal 14-740,07000 Ciudad de M\'exico, Mexico}

\authorrunning{Angel S. Miramontes et al.}
%
\date{Received: date / Revised version: date}
%
\abstract{
We investigate the properties of the radially excited charged pion, with a specific focus on its electromagnetic form factor (EFF) and its box contribution to the hadronic light-by-light (HLbL) component of the muon's anomalous magnetic moment, $a_{\mu}$. Utilizing a coupled non-perturbative framework combining Schwinger-Dyson and Bethe-Salpeter equations, we first compute the mass and weak decay constant of the pion's first radial excitation. Initial results are provided for the Rainbow-Ladder (RL) approximation, followed by an extended beyond RL (BRL) analysis that incorporates meson cloud effects. Building on our previous work, this analysis demonstrates that an accurate description of the first radial excitation can be achieved without the need for a reparametrization of the interaction kernels. Having demonstrated the effectiveness of the truncation scheme, we proceed to calculate the corresponding EFF, from which we derive the contribution of the pion's first radial excitation to the HLbL component of the muon's anomalous magnetic moment, producing  $a_{\mu}^{\pi_1-\text{box}}(\text{RL}) = -(2.03 \pm 0.12) \times 10 ^{-13}$, 
 $a_{\mu}^{\pi_1-\text{box}}(\text{BRL}) = -(2.02 \pm 0.10) \times 10 ^{-13}$. Our computation also sets the groundwork for calculating related pole contributions of excited pseudoscalar mesons to $a_{\mu}$. }

%
\titlerunning{Radially excited pion: EFF and box contribution to the muon's $g-2$}
\maketitle

\section{Introduction}
Understanding the origin of the mass spectrum and the structural properties of hadrons is, unquestionably, one of the key challenges in modern physics. Nucleons (protons and neutrons) naturally take a center stage in this endeavor, being the foundations of atomic nuclei and thus contributing predominantly to the visible matter\,\cite{Proceedings:2020fyd,Aznauryan:2012ba}. At the same time, the importance of providing a simultaneous description of the pion and kaon, the lightest hadrons in nature, is becoming increasingly clear\,\cite{Raya:2024ejx,Ding:2022ows,Roberts:2021nhw}: at a certain level of approximation, these systems are the messengers of the nuclear force\,\cite{Yukawa:1935xg}; and, on the other hand, their origin is completely linked to the emergence of mass within the Standard Model, as they are regarded as the Nambu-Goldstone (NG) bosons associated with dynamical chiral symmetry breaking (DCSB)\,\cite{Horn:2016rip}. The interest in $\pi-K$ mesons is further boosted by advances in modern experimental facilities, which enable their structure to be mapped with unprecedented precision\,\cite{Accardi:2023chb,Quintans:2022utc,Anderle:2021wcy,BESIII:2020nme,Accardi:2012qut}. From a theoretical point of view, we expect quantum chromodynamics (QCD), one of the fundamental forces defining the Standard Model (SM) of Salam, Weinberg and Glashow, to be the underlying theory describing  strong nuclear interactions\,\cite{Marciano:1977su,Marciano:1979wa}. However, progress in unraveling QCD presents a multitude of challenges. Firstly, the fundamental degrees-of-freedom, \emph{i.e.} quarks and gluons, cannot be observed in isolation; instead, due to color confinement, the detectable objects are color-singlet bound-states known as hadrons. Secondly, while asymptotic freedom causes the strong interaction coupling to diminish at high energy scales, making a perturbative approach to QCD possible, several static and dynamic properties of hadrons are dictated by the opposite end of the energy domain\,\cite{Roberts:2020hiw}. Therefore, the infrared complexities of QCD demand the use of non-perturbative methods, such as lattice QCD\,\cite{Aoki:2016frl,Edwards:2011jj}, the Schwinger-Dyson equations (SDE) approach\,\cite{Eichmann:2016yit,Roberts:1994dr}, and effective theories\,\cite{Ecker:1988te
,Roig:2019reh,Estrada:2024cfy}.

A fundamental aspect of grasping and tackling the challenges of the strong interactions is the characterization of the excited states of hadrons\,\cite{Aznauryan:2012ba,Ramalho:2023hqd}. This understanding helps us identify how modifications in these excited systems emerge, although these are composed of the same constituents as the corresponding ground states\,\cite{Carman:2023zke,Paredes-Torres:2024mnz}. With that in mind, the present study focuses on studying aspects of the static and structural properties of the pion's first radial excitation, $\pi(1300)$, including those revealed by its elastic electromagnetic form factor (EFF). Certain characteristics of the $\pi(1300)$ are anticipated whatsoever:
\begin{itemize}
    \item As a first radial excitation, the $\pi(1300)$ system is expected to exhibit a node in its corresponding wavefunction\,\cite{Holl:2004fr,Li:2016dzv}.
    \item The mass of the ground-state pion ($m_\pi=0.139$ GeV) is roughly $1/5$ that of its vector meson counterpart, the $\rho$ meson. However, for the first excitations, this ratio is approximately $\sim 0.9$,\,\cite{ParticleDataGroup:2024cfk}.
    \item In contrast to the NG modes, which posses a non-zero leptonic decay constant in the chiral limit ($f_\pi^0\approx 0.130$ GeV), the decay constants of their radial excitations must vanish\,\cite{Dominguez:1977nt,Dominguez:1976ut}. For physical quark masses, these should remain small, $\lesssim 0.1 \,f_\pi^0$\,\cite{Xu:2022kng,Maltman:2001gc,Andrianov:1998kj,Diehl:2001xe}.
\end{itemize}
Throughout this work, we employ the coupled formalism of SDEs and the Bethe-Salpeter Equation (BSE) for two-particle relativistic bound-states. This approach has become an increasingly powerful non-perturbative tool for studying hadron physics, enabling a wide variety of their properties to be addressed, see \emph{e.g.}\,\cite{Chang:2013pq,Chang:2013nia,Raya:2015gva,Ding:2019qlr,Raya:2019dnh,Cui:2020tdf,Xu:2023izo,Yao:2024ixu,Miramontes:2021xgn,Miramontes:2022uyi,Miramontes:2021exi,Raya:2022ued}. The SDE-BSE framework captures essential traits of QCD, such as confinement and DCSB; it allows hadron-related observables to be largely traced back to the theory's Green functions, thereby preserving a direct link to QCD. We expect and observe this link to persist for the excited states such as $\pi(1300)$. Regarding the latter, we focus on the computation of its space-like EFF. For this purpose, we employ both the typical Rainbow-Ladder (RL) approximation\,\cite{Munczek:1994zz,Bender:1996bb}, and a beyond RL (BRL) scheme that permits us to incorporate meson cloud effect (MCE) in a rigorous manner,\,\cite{Miramontes:2021xgn,Miramontes:2022uyi,Miramontes:2021exi,Miramontes:2019mco}. The MCE is anticipated to be crucial in the time-like EFFs. Its contribution in the space-like domain is nuanced and limited to $Q^2 \gtrsim 0$\,\cite{Miramontes:2021xgn,Miramontes:2022uyi,Alkofer:1993gu}. However, this is precisely the domain that determines the corresponding box-diagram hadronic light-by-light (HLbL) contribution to the muon anomalous magnetic moment, $a_\mu=(g_\mu-2)/2$,\,\cite{Miramontes:2021exi,Raya:2022ued}. On the other hand, the remarkable success of the SM in explaining visible matter continues to be unrattled. Only precision observables, such as $a_\mu$, can put it to test\,\cite{Aoyama:2020ynm,Colangelo:2022jxc}. For these reasons, extending our previous explorations on the ground-state $\pi-K$ box\,\cite{Miramontes:2021exi} and pseudoscalar-pole contributions\,\cite{Raya:2019dnh}, we  also evaluate the corresponding contribution to the anomalous magnetic moment of the muon, {\em i.e.}, $a_\mu^{\pi(1300)-\text{box}}$. The produced outcomes shall be contrasted with expectations from vector meson dominance (VMD). Likewise, it will be analyzed to what extent this result is influenced by the details of the corresponding Bethe-Salpeter wavefunction.

The manuscript is organized as follows. In Section II, we recall the key ingredients of the SDE/BSE formalism and in Section III the truncations employed. Section IV presents numerical results for the EFF of the pion's first radial excitation $\pi(1300)$ for space-like photons. We also outline the essential components required to analyze its box-diagram HLbL contribution to $a_{\mu}$ and proceed to compute them explicitly. Finally, in Section V, we summarize our conclusions and discuss the potential avenues for future research.

\section{SDE and BSE formalism} \label{SDEBSE}
In this section we summarize the main elements of the non-perturbative approach to QCD, based on the SDEs and BSEs, to compute the $\pi(1300)$ space-like EFF. For a more comprehensive and pedagogical review of the topic, we refer the interested reader to Refs.~\cite{Eichmann:2016yit,Roberts:1994dr}. It is important to note that all calculations are carried out in the Euclidean space-time. \\

\noindent {\bf{Quark propagator:}}
The quark propagator can be obtained by solving its SDE, which takes into account the self-interactions of quarks through their interaction with the gluon field. The dressed quark SDE reads as follows:
\begin{align}\label{eq:DSE}
	S^{-1}_f(p) =& Z_2 \,S^{-1}_{0,f}(p) - Z_{1f} \,g^2 \,C_F  \nonumber\\
	&\times\int \!\! \frac{d^4q}{(2\pi)^4}\, i\gamma^\mu S_f(q) \,\Gamma_{\mathrm{qg},f}^\nu(q,p) \,D^{\mu\nu}(k)\,,
\end{align}
where $f$ denotes the flavor of the quark. The renormalization constants $Z_{1f}$ and $Z_2$ account for the renormalization of the quark-gluon vertex and the quark propagator, respectively. 
Additionally, the constant $C_F = \frac{4}{3}$ corresponds to the color Casimir, considering the number of colors in the fundamental representation as $N_c = 3$.
The inverse tree-level propagator  is given by
\begin{equation}
     S^{-1}_{0,f}(p) = i\slashed{p}+Z_m \,m_f \,,
\end{equation}
where $m_f$ represents the renormalized (by the renormalization factor $Z_m$) quark mass obtained from the QCD action. The dressed quark-gluon vertex, $\Gamma_{\mathrm{qg},f}$, encodes a complex structure that incorporates not only gluonic interactions but also effective meson exchange contributions. The symbol $D_{\mu \nu}(k)$ represents the full gluon propagator. In the Landau gauge, it can be expressed as follows:
\begin{equation}
D_{\mu \nu}(k) = \left( \delta_{\mu \nu} - \frac{k_\mu k_\nu}{k^2}\right) \frac{Z(k^2)}{k^2}~,
\end{equation}
with  $Z(
k^2)$ being the gluon dressing function. For the simplicity of notation, we have suppressed the color indices. The solution for the dressed quark propagator, Eq.~\eqref{eq:DSE}, can be written in the form, 
\begin{align}\label{QuarkProp}
S_f(p) = \frac 1 {A_f(p^2)} \frac {-i \slashed{p} +M_f(p^2)}{p^2+M^2_f(p^2)}\,.
\end{align}
For each quark flavor $f$, the dressed quark propagator is described by two independent dressing functions, namely $A_f(p^2)$ and $M_f(p^2)$. The latter represents the dynamically generated mass function. It can be identified with the running mass of the quark as a function of momentum squared $p^2$. These dressing functions provide a comprehensive description, not only of the quark propagator, but also of the three-point interaction vertex between quarks and gluons, incorporating quantum corrections as well as nonperturbative aspects of QCD.\\

\noindent {\bf{The Bethe-Salpeter Amplitudes:}}
Mesons as relativistic bound states of a quark and an anti-quark are described by the Bethe-Salpeter amplitude (BSA), $\Gamma(p,P)$. It can be derived via the homogeneous BSE, which reads,
\begin{eqnarray}
\label{eq:homogeneousBSE}
\bracket{\Gamma}_{a\alpha,b\beta}\bracket{p,P}= \int_q K^{r\rho,s\sigma}_{a\alpha,b\beta}\bracket{P,p,q} \times S_{r\rho,e\epsilon}\bracket{k_1} \nonumber\\
\bracket{\Gamma}_{e\epsilon,n\nu}\bracket{q,P}S_{n\nu,s\sigma}\bracket{k_2} \,.
\end{eqnarray}
The total momentum of the meson is represented by $P$, the relative momentum between the quark and antiquark is denoted by $p$, whereas the internal relative momentum, which gets integrated over, is denoted as $q$ (note that the symbol $\int_q := \int \frac{d^4q}{(2\pi)^4}$ stands for a Poincar\'e covariant 4-momentum integration). The internal quark and antiquark momenta are conveniently defined as $k_1=q+P/2$ and $k_2=q-P/2$, respectively, such that $P=k_1-k_2$ and $q=(k_1+k_2)/2$. Latin  letters represent Dirac indices while the Greek letters represent flavour indices. The Dirac part of the BSA can be expanded in a tensor basis which, in the case of pseudoscalar mesons, consists of four independent tensor structures\,\cite{Maris:1997hd}. These basis tensors provide a suitable framework to describe the spatial as well as the spin structure of the meson. The two-body interaction kernel $K$ encodes all possible interactions between the quarks and antiquarks within the bound state. It naturally takes into account the strong interaction dynamics and it contributes to the overall structure and properties of the meson under consideration. As detailed later on, the specific form of the two-body kernel would be linked to the truncation scheme applied in the SDE of the quark propagator.

It is important to note that solving Eq.\,\eqref{eq:homogeneousBSE} provides the bound-state mass and the corresponding BSA. Once the BSA is properly normalized,\,\cite{Nakanishi:1965zza}, the pseudoscalar meson leptonic decay constant follows directly from the well-known expression~\cite{Maris:1997hd}, 
\begin{eqnarray}
    f_{\textbf{P}} m^2_{\textbf{P}} = \sqrt{N_c} Z_2 \text{Tr} \int \frac{d^4q}{(2\pi)^4} \gamma^5 \slashed{P} S(k_1) \Gamma\bracket{q,P}S(k_2)\,.
\end{eqnarray}
Both mass and decay constants strictly constrain the involved model parameters, which allows these to be fixed with great precision. This will be addressed later. \\

\noindent {\bf{Quark photon-vertex:}}
The quark-photon interaction vertex (QPV) is a quantity of particular interest in studying the electromagnetic interactions of quarks. It describes the coupling between quarks and photons and plays a crucial role in processes involving electromagnetic probes.  The fully-dressed QPV, denoted as $\Gamma^{\mu}$, can be described using an inhomogeneous BSE as follows:
\begin{eqnarray}
\label{eq:inhomBSE_vector}
\bracket{\Gamma^{\mu}}_{a\alpha,b\beta}\bracket{p,Q}&=&
Z_2 \bracket{\gamma^\mu}_{ab} t_{\alpha\beta}\\
&+&\int_q K^{r\rho,s\sigma}_{a\alpha,b\beta}\bracket{Q,p,q} \, S_{r\rho,e\epsilon}\bracket{k_1}
\nonumber \\
&\times& \bracket{\Gamma^{i,\mu}}_{e\epsilon,n\nu}\bracket{Q,q}S_{n\nu,s\sigma}\bracket{k_2} \,,
\nonumber
\end{eqnarray}
where the symbols have the usual meaning already explained. As far as the kinematics are concerned, $Q$ denotes the probing photon momentum, while $p$ and $q$ represent the external and internal relative momenta between the quark and antiquark, respectively. The internal quark and antiquark momenta are defined as $k_1=q+Q/2$ and $k_2=q-Q/2$, ensuring that $Q=k_1-k_2$ and $q=(k_1+k_2)/2$. Again, the Latin letters are used to denote Dirac indices, while the Greek letters represent flavour indices. The isospin structure of the vertex is given by $t_{\alpha\beta} = \textrm{diag}\bracket{\nicefrac{2}{3},\nicefrac{-1}{3}, \nicefrac{-1}{3}}$. Our calculation includes all eight basis vectors transverse to the photon momentum and four non-transverse vectors, thus constituting a complete basis for the decomposition of the QPV\,\cite{Albino:2018ncl,Bermudez:2017bpx}.\\

\noindent {\bf{Electromagnetic Form Factor:}}
The interaction between a virtual photon and a pseudoscalar meson can be described by a single EFF, $F_M(Q^2)$, which can be conveniently expressed in terms of the matrix element of the electromagnetic current as:
\begin{equation}
\label{eq:FF}
\langle\textbf{P}(p_1)|J^\mu|\textbf{P}(p_2)\rangle= e(p_1 + p_2)^{\mu} F_M(Q^2)\,.
\end{equation}
Here, $Q=p_1 -p_2$ represents the four-momentum of the probing photon, $e$ is the elementary electromagnetic charge, $\textbf{P}(p_1)$ and $\textbf{P}(p_2)$ denote the incoming and outgoing meson states, respectively. On the other hand, at the level of the meson BSA, the dressed quark propagators and the strong interactions, this electromagnetic current $J^{\mu}$ can be written as follows
\begin{equation}
J^{\mu} = \bar{\Psi}_{\textbf{P}}^f G_0 (\mathbf{\Gamma}^{\mu} - K^{\mu}) G_0 \Psi_{\textbf{P}}^i ~.
\label{eq:Current}
\end{equation}
In this expression, $\Psi_{\textbf{P}}^{i}$ and  $\Psi_{\textbf{P}}^{f}$ represent the BSA of the incoming and outgoing meson, respectively. $G_0$ encodes the appropriate product of the dressed quark propagators. Moreover, $\mathbf{\Gamma}^{\mu}$ corresponds to the impulse approximation (IA) diagrams, representing only the coupling of photons to the dressed valence quarks. It is expressed as follows:
\begin{equation}
\mathbf{\Gamma}^{\mu} = \left(S^{-1} \otimes S^{-1} \right)^{\mu} = \Gamma^{\mu}\otimes S^{-1} + S^{-1}\otimes \Gamma^{\mu} \,,
\end{equation}
where $S^{-1}$ is the inverse quark propagator. The second term in expression~\eqref{eq:Current}, involving $K^\mu$, accounts for the effects beyond the impulse approximation and represents the interaction of the photon with the Bethe-Salpeter interaction kernel. Inclusion of both terms in the electromagnetic current, Eq.~\eqref{eq:Current}, is essential in ensuring electromagnetic charge is conserved.

In order to compute the  electromagnetic current in Eq.~\eqref{eq:Current}, it is necessary to carefully analyze the ingredients involved, such as the quark propagators, meson BSA, the three-point QPV and the corresponding interaction kernels. This is what we set out to do in the next section.

\section{Truncation} \label{Truncations}
To solve the infinitely coupled system of integral equations, non-perturbative truncations are required for the interaction kernel of the BSE together with the SDE for the quark propagator\,\cite{Qin:2020jig,Binosi:2016rxz,Lessa:2022wqc}. These truncations must adhere to symmetry principles, as well as the matching with the perturbation theory in the weak coupling regime~\cite{Bermudez:2017bpx,Guzman:2023hzq,Sultan:2018tet,Bashir:2011dp}. One important aspect is the correct implementation of chiral physics which ensures that ground-state pions become massless bound states when the current quark masses are set to zero, the chiral limit\,\cite{Munczek:1994zz,Bender:1996bb}; in other words, the emergence of pions as NG modes must be guaranteed. Additionally, the $U(1)$ vector symmetry entails charge conservation, which is crucial for the accurate calculations of EFFs. To achieve these objectives, the truncation must satisfy two important identities: the Axial Vector Ward-Green-Takahashi identity (AxGWTI) for appropriate implementation of chiral symmetry, and the Vector Ward-Green-Takahashi identity (VWGTI) for correct incorporation of charge conservation\,\cite{Qin:2013mta}.

In order to calculate the quark propagator, the BSA, and the EFF of the $\pi(1300)$, we employ a truncation that includes (i) a flavor-blind dressed quark-antiquark gluon exchange that supplies the necessary interaction strength for forming mesonic bound states, and
(ii) a meson-exchange mechanism that serves as a faithful representation of the MCE.

\subsection{Rainbow-ladder and Meson cloud}
A straightforward non-perturbative truncation that simultaneously preserves the AxWGTI and the VWGTI within the SDE/BSE formalism is known as the RL truncation. In this truncation, the BSE interaction kernel is simplified to a vector-vector gluon exchange with an effective coupling denoted as $\alpha(k^2)$. The interaction kernel can be expressed as\ follows,:
\begin{equation}
K^{r\rho,s\sigma}_{a\alpha,b\beta}\bracket{Q,p,q}=\alpha\bracket{k^2}\gamma^\mu_{ar}\gamma^\nu_{sb}D_{\mu\nu}\bracket{k}\delta^{\alpha\rho}\delta^{\sigma\beta}~,\label{eq:RLkernel}
\end{equation}
where we have omitted the color indices for the sake of notational simplicity.
Here, $k=p-q$ represents the momentum flowing through the gluon propagator. The effective coupling $\alpha(k^2)$ determines the strength of the quark-antiquark interaction in this truncation.
Furthermore, the structure of the truncated quark SDE is driven by the corresponding consistent substitution:
\begin{equation}
Z_{1f} \gamma_{\mu} Z(k^2) \Gamma_{\nu}^{\text{qgl}}(q,p) \rightarrow Z_2^2 \gamma_{\mu} 4\pi \alpha(k^2) \gamma_{\nu} \,.
\label{eq:RL_SDE}
\end{equation}
 These substitutions corresponds to the RL truncation.

For the purpose of this article, we employ the widely and reliably used Maris-Tandy (MT) interaction model to mimic the effective coupling both in the perturbative and the non-perturbative regions. It is described by the following expression~\cite{Maris:1997tm}:
\begin{eqnarray}
\alpha(q^2) &=&
 \pi\eta^7\left(\frac{q^2}{\Lambda^2}\right)^{2}
e^{-\eta^2\frac{q^2}{\Lambda^2}} \nonumber \\
&+&\frac{2\pi\gamma_m
\big(1-e^{-q^2/\Lambda_{t}^2}\big)}{\textnormal{ln}[e^2-1+(1+q^2/\Lambda_{QCD}
^2)^2]} \,. \label{eq:MTmodel}
\end{eqnarray}
This truncation model of the SDEs is composed of two terms. The first one, which contains a Gaussian function, dominates in the infrared region and provides sufficient interaction strength for the right amount of DCSB to occur. Note that the specific functional form of this term in the deep infrared has no significant effect on the results, as long as it provides adequate infrared enhancement\,\cite{Sultan:2018tet}. The second term dominates in the ultraviolet region and reproduces the one-loop behavior of the quark propagator in QCD at large momenta. The MT model includes two free parameters, namely $\Lambda$ and $\eta$, which are typically determined by fitting the model to match the ground-state pseudoscalar mass and its weak decay constant. In the SDE/BSE framework, the running quark masses $m_u$ and $m_d$ are also introduced as input parameters. Additionally, we incorporate a scale $\Lambda_t=1$ GeV, which is included for technical reasons and does not affect the computed observables. The anomalous dimension $\gamma_m$ is given by $\gamma_m=12/(11N_c-2N_f)=12/25$, where $N_f=4$ represents the number of quark flavors and $N_c=3$ as mentioned before. Moreover, we adopt the QCD mass scale of $\Lambda_{QCD}=0.234$ GeV.

At this point, it is worth pointing out that the RL truncation is generally dependable for most pion, kaon, and nucleon related observables, primarily due to its preservation of key symmetries (see \emph{e.g.} Refs.\,\cite{Chang:2013nia,Raya:2015gva,Ding:2019qlr,Raya:2019dnh,Cui:2020tdf,Xu:2023izo,Yao:2024ixu}). However, when applied to different systems, including the excited states, it requires improvement. For instance, it has been shown in the literature, including in Refs.\,\cite{Holl:2004fr,Xu:2022kng,Rojas:2014aka}, that the simplicity of the RL truncation does not produce an excited pion mass that is sufficiently large. Typically, this kind of problems are circumvented by artificially inflating the infrared parameters of the interaction kernel, which is rather unsatisfactory\,\cite{Qin:2011dd}. Other hadron-related quantities further highlight the limitations of the RL truncation. As far as EFFs are concerned, the RL scheme turns out to be insufficient for calculating time-like EFF, as it does not properly account for the analytic structure required in this regime, since it treats meson resonances as stable particles, leading to real-valued masses without any decay widths\,\cite{Miramontes:2021xgn,Miramontes:2022uyi}. 

With the above discussion in mind, following our previous investigations~\cite{Miramontes:2021xgn,Miramontes:2021exi,Miramontes:2019mco}, we shall supplement the gluon exchange term between the quark and the antiquark with the inclusion of explicit mesonic contributions into the SDE/BSE system. These ideas were introduced to this formalism, firstly, in  Refs.~\cite{Fischer:2007ze,Sanchis-Alepuz:2014wea,Fischer:2008sp}. This extended scheme is the BRL truncation. Technical details on the BRL truncation can be found in Appendix A and in Refs.~\cite{Miramontes:2021xgn,Miramontes:2021exi,Miramontes:2019mco}. The inclusion of meson exchange contributions in the BRL truncation introduces a multi-particle branch cut in the quark-photon vertex, starting at the two-pion production threshold. This modification leads to a form factor with the correct analytic structure in the time-like region, addressing the unphysical results observed in simpler truncations. As a result, the BRL truncation enables theoretical predictions that align more closely with experimental observations, demonstrating its improved accuracy over traditional approaches.

\section{Numerical solutions} \label{Solutions}

We divide this section into three natural subsections. First, the masses, decay constants, and BSAs of the pion's first radial excitation are computed using both RL and BRL truncations in the process. After obtaining the best description of the aforementioned quantities, we compute the corresponding EFF; this is done along the same lines we do it for the ground-state pion\,\cite{Miramontes:2021exi}. By employing the well-known master formula\,\cite{Colangelo:2017fiz}, we later use the box diagram to calculate the radial pion's HLBL contribution to the anomalous magnetic moment of the muon.   

\begin{figure}[t]
\centerline{%
\includegraphics[width=1.0\columnwidth]{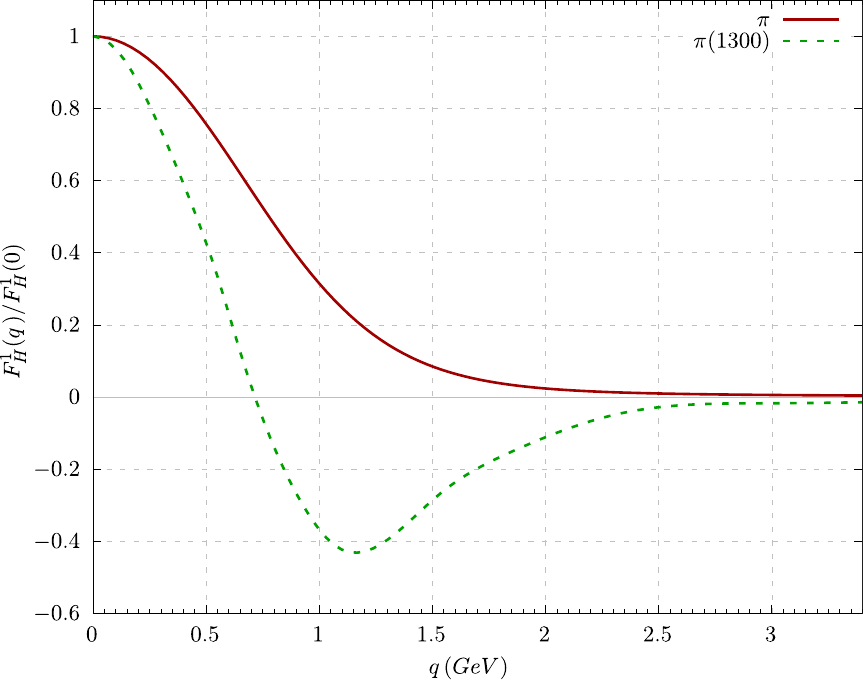}}
\caption{Leading Chebyshev moment of the dominant BSA: $H=\pi$, the ground-state pion (solid), and $H=\pi(1300)$, its first radial excitation (dashed). }
\label{fig:BSA}     
\end{figure}

\subsection{Radial excitation } \label{Radial_excitation}
The homogeneous BSE, Eq.~\eqref{eq:homogeneousBSE} (either with the RL or the BRL truncation) can be solved numerically by transforming this equation into an eigenvalue problem. This is plainly achieved by introducing a function $\lambda(P^2)$ on the right hand side of the BSE. The physical solutions are located on the mass shell points $P^2_n = -M^2_n$, and occur when $\lambda(P^2_n=-M^2_n)=1$; here $M^2_0$ corresponds to the ground state mass of the meson and $M^2_n$ ($n \geq 1$) is then the $n^{\rm th}$ radial excitation mass squared. The calculation of the meson BSAs can be simplified when we expand them into Chebyshev polynomials of the second kind, where the angular dependence can be factor
ed out (consult Appendix~\ref{Cheby}). For a detailed discussion, we recommend referring to Ref.~\cite{Sanchis-Alepuz:2017jjd}. Additionally, pseudoscalar mesons with the same quark masses are $C$-parity eigenstates, which entails that the meson BSAs are even in the angular variable $z=q\cdot P$. Therefore, when the BSAs are factorized into Chebyshev polynomials the only contributions arise from the even Chebyshev moments (see e.g. Ref.~\cite{Holl:2004fr}). A total of six Chebyshev polynomials are sufficient for a proper description of the pion excited state.

On another relevant numerical aspect, it is well known that in order to solve the corresponding homogeneous BSE, the quark SDE has to be sampled in the complex plane. In this article, the analytical continuation to the complex plane is performed via the Cauchy theorem (see Appendix~\ref{Cauchy}). Nevertheless, the numerically accessible region is limited by the position
of the first pair of complex-conjugate poles of the quark
propagator. For this reason, we use a parameterization of the quark propagator in terms of pairs of complex conjugate poles fitting the solution in the complex plane. Further details on this issue can be found in Appendix~\ref{Cauchy}.

Notably, the radial excitations were calculated by fixing the free parameters of the MT interaction to $\Lambda = 0.78$, $\eta \in \{1.60,1.65,1.70\}$ and $m_{u/d} = 3.7$ MeV - the same used for the ground-state pion. The resulting masses and decay constants are collected and compared to measurements in Table~\ref{tab:masses_and_decay_constants}. As can be seen, the mass of the $\pi(1300)$ is perfectly aligned with its empirical value\,\cite{ParticleDataGroup:2024cfk}. The decay constant is small, and it has the expected order of magnitude\,\cite{Xu:2022kng,Maltman:2001gc,Andrianov:1998kj,Diehl:2001xe} and, in fact, matches the expectations from QCD sum rules, $-f_{\pi(1300)}=1.6(3)$ MeV\,\cite{Maltman:2001gc}. Moreover, in the chiral limit, the latter would be identically zero, thus fulfilling the requirements of symmetry principles\,\cite{Dominguez:1977nt,Dominguez:1976ut}. This significant result demonstrates that the BRL truncation adheres to the requirements of the AxWGTI\,\cite{Li:2016dzv}. Concerning the structure of the BSA, Fig.\,\ref{fig:BSA} reveals that the $0-$th Chebyshev moment of the dominant amplitude (that attached to the $\gamma_5$ structure,\,\cite{Maris:1997hd}), develops a node.  The zero-crossing occurs at $k^2\approx0.6\,\text{GeV}^2$. It reflects the excited-state nature of the $\pi(1300)$ and highlights the structural differences compared to the ground state. These distinctions, among other factors, are also evident in the distribution amplitudes\,\cite{Li:2016dzv}.

As previously discussed, the RL truncation is limited in its ability to describe both the ground state and the first radial excitation of the pion simultaneously with a single set of parameters in the MT interaction. Specifically in our case, while the RL truncation can yield a good estimate of the $\pi(1300)$ mass, it fails to accurately capture the decay constant of the ground state, leading to a marked deviation from expected values.

In contrast, our findings demonstrate that the BRL truncation effectively overcomes the limitations of the RL approach, in exchange for an \emph{a priori} minimal violation of the VWGTI. This issue is resolved as in Appendix A. On the flip side, by incorporating meson exchange contributions, the BRL truncation allows for a consistent and accurate description of both the ground state and the first radial excitation of the pion using a single, unified set of parameters within the MT interaction. This not only simplifies the modeling process but also enhances the predictive power of the framework, as it avoids the need for separate parameter tuning for different states. This outcome highlights the strength of the BRL truncation in accurately capturing the properties of both states within a unified framework, offering an improvement over the RL truncation for the first radial excitation of the pion.

With the accurate description of both states at hand, the BRL truncation can potentially lead to a good description of the EFF of the $\pi(1300)$ and, subsequently, its box contribution to the muon $g-2$. 

\begin{table}
    \centering
    \sisetup{table-format=1.3(2), table-number-alignment=center}
    \rowcolors{2}{gray!25}{white}
    \begin{tabular}{l *{4}{S}}
        \toprule
        \rowcolor{gray!30}
        & {$m_{\pi}$} & {$f_{\pi}$} & {$m_{\pi(1300)}$} & {$f_{\pi_1}$} \\
        \midrule
        RL   & 0.145(2) & 0.139(2) & 1.32(4) & -0.0019(4) \\
        BRL  & 0.139(2)   & 0.130(2) & 1.29(4) & -0.0017(4) \\
        Exp. & 0.139   & 0.130 & 1.30 & \\
        \bottomrule 
    \end{tabular}
\\    \caption{Masses and decay constants computed in RL and BRL truncations compared with the experimental (Exp.) data, extracted from~\cite{10.1093/ptep/ptac097}. The blank space indicates that there is no data available. All values are given in GeV.} 
    \label{tab:masses_and_decay_constants}
\end{table}

\begin{figure*}[t]
\centerline{%
\includegraphics[width=0.8\textwidth]{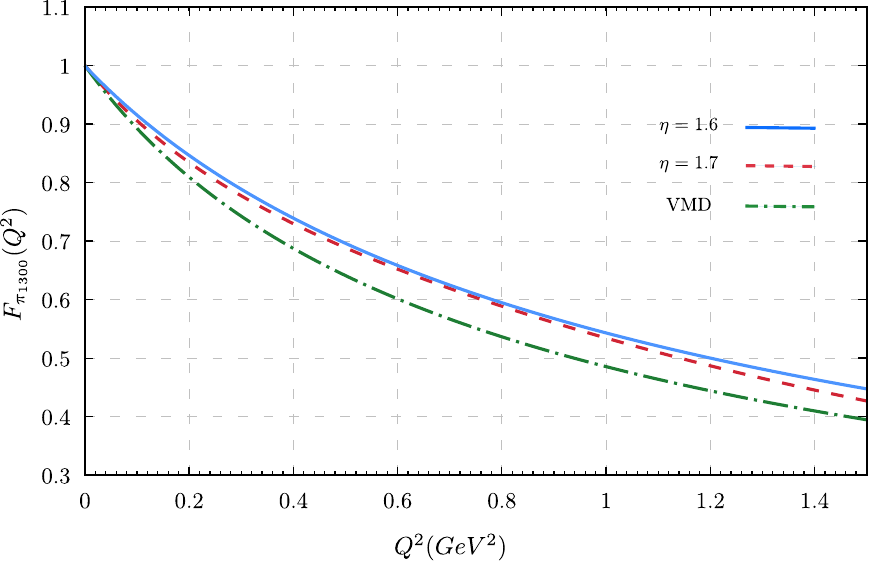}}
\caption{The $\pi$(1300) EFF for a space-like momentum $Q^2$ calculated with BRL truncation for two values of $\eta$ and compared with the VMD model from Eq.~(\ref{eq:VMD}).}
\label{fig:EFF}     
\end{figure*}

\subsection{$\pi(1300)$ electromagnetic form factor}

Building on the framework established in the previous sections, we proceed to calculate the EFF of the $\pi(1300)$ using the same set of free parameters that successfully describe both the ground state and the first radial excitation of the pion simultaneously. The EFF is computed by evaluating the electromagnetic current as defined in Eq.\,\eqref{eq:Current}.
To manage the computational complexity, the EFF is calculated within the impulse approximation\,\cite{Chang:2013nia}. While this approach simplifies the numerical calculations, it still captures the essential features of the $\pi(1300)$'s electromagnetic structure. The final outcome is presented in Fig.~\ref{fig:EFF}.
The corresponding charge radius follows from the standard definition:
\begin{equation}
    \label{eq:ChargeRadii}
    r^2 = -6 \frac{dF(Q^2)}{dQ^2}\big|_{Q^2=0}\;;
\end{equation}
by applying this equation, we obtain a charge radius for the first radial excitation of the pion, $r_{\pi(1300)} = (0.583 \pm 0.010)~\mathrm{fm}$; the uncertainty here accounts for the variation of $\eta$ within the selected range. Notably, this value is comparable to that of the ground state, with $r_{\pi(1300)}/r_{\pi}\approx 0.88$, despite the significantly larger mass of the excited state, $m_{\pi(1300)}/m_{\pi}\approx 9.3$. Among other implications, it suggests that the spatial extent of the excited state and the ground state are comparable. In order to compare our results with other methods, we also estimate the $\pi(1300)$ EFF using a vector meson dominance (VMD) model. The VMD approach offers a phenomenological description of the form factor, where the photon is assumed to couple to the pion through intermediate vector mesons. The form factor is modeled as\,:
\begin{eqnarray}
    F_{\pi^{\prime}} (Q^2) = \Bigg(\frac{m_{\rho}^2}{m_{\rho}^2 + Q^2} + \beta \frac{m_{\rho^{\prime}}^2}{m_{\rho^{\prime}}^2 + Q^2} \Bigg)/(1+ \beta),
    \label{eq:VMD}
\end{eqnarray}
with $\beta = 0.36 \pm 0.10$, (estimated from the parametrization in Ref.~\cite{CMD-3:2023alj} for $F_\pi(Q^2)$). Here, $m_{\rho}$ and $m_{\rho^{\prime}}$ are the masses of $\rho$ and $\rho^{\prime}$ mesons, respectively, and $Q^2$ is the probing photon momentum squared. The VMD model offers a complementary approach to our direct calculation within the BRL truncation, providing a useful cross-check for the accuracy and consistency of our explorations. The VMD computation is adequate as long as the first pion and kaon excitations have the same flavor structure as the ground states, which is precisely our case. The VMD-derived EFF, calculated using Eq.(\ref{eq:VMD}), is displayed alongside our primary results in Fig.\ref{fig:EFF}, allowing for a direct comparison between the two methods.

\subsection{Box contributions to HLbL}

Following our previous effort~\cite{Miramontes:2021exi}, we investigate the $\textbf{P}$-box  contributions ($\textbf{P}=\pi(1300)$) denoted as $a_{\mu}^{\textbf{P}-box}$. To calculate such 
 hadronic light by light (HLbL) contribution to the muon's anomalous magnetic moment, $a_{\mu}$,
 we employ the master formula derived in~\cite{Colangelo:2017fiz}, which reads:
\begin{equation}
a_{\mu}^{\textbf{P}-box} = \frac{\alpha^3_{em}}{432 \pi^2} \int_{\Omega}  \sum_i^{12} T_i(Q_1,Q_2,\tau) \bar{\Pi}_i^{\textbf{P}-box}  (Q_1,Q_2,\tau),
\label{eq.master_formula}
\end{equation}
where $\alpha_{em}$ is the QED coupling constant and $\int_{\Omega}$ denotes the integration over the photon momenta, $Q_{1,2}$, and the cosine of their relative angle, $\tau$. With $Q_3^2=Q_1^2+Q_2^2+2|Q_1|  |Q_2|\tau$, the functions $\bar{\Pi}_i^{\textbf{P}-box}$ are expressed as:
\begin{eqnarray}
\label{I_feynman}
\bar{\Pi}_i^{\textbf{P}-box}(Q_1^2,Q_2^2,Q_3^2) &=& F_{\textbf{P}}(Q_1^2)F_{\textbf{P}}(Q_2^2) F_{\textbf{P}}(Q_3^2)\\
&\times& \frac{1}{16 \pi^2} \int_0^1 dx \int_0^{1-x} dy I_i(x,y)\,;\nonumber
\end{eqnarray}
the scalar functions $T_i$ and $I_i$ are provided in Appendices B and C, respectively, of  Ref.~\cite{Colangelo:2017fiz}. Thus, the only ingredient we need is the $\pi(1300)$ EFF, $F_{\textbf{P}}(Q^2)$ which we have evaluated in the previous sub-section.

After determining the $\pi(1300)$ EFF, we can compute its box contribution to $a_{\mu}$ from the master formula of Eq.~(\ref{eq.master_formula}), where the integration can be performed by employing the CUBA integration library~\cite{Hahn:2004fe}. First, using the VMD model of Eq.~(\ref{eq:VMD}), we obtain the following contribution,
\begin{equation}
    a_{\mu}^{\pi(1300)-\text{box-VMD}} = -(1.85 \pm 0.06) \times 10 ^{-13} ,
\end{equation}
where we have used the values $m_{\rho}=0.775$ GeV, $m_{\rho^{\prime}}=1.465$ GeV and $m_{\pi^{\prime}} =1.30$ GeV.
On the other hand, numerically computing the EFF by employing the RL and BRL truncations, we get the following estimates,
\begin{eqnarray}
    a_{\mu}^{\pi(1300)-\text{box}}(\text{BRL}) &=& -(2.02 \pm 0.10) \times 10^{-13}. \\ 
    a_{\mu}^{\pi(1300)-\text{box}}(\text{RL}) &=& -(2.03 \pm 0.12) \times 10^{-13}. 
\end{eqnarray}
Clearly, this contribution is subdominant compared with of the ground-state pion. For instance, adopting the BRL produced value, and the pion ground-state box contribution estimated in~\cite{Miramontes:2021exi},
$a_{\mu}^{\pi-\text{box}} = -1.56(2)\times 10^{-10}$, one finds the following ratio:
\begin{equation}
    \frac{a_{\mu}^{\pi(1300)-\text{box}}}{a_{\mu}^{\pi-\text{box}}}=1.29(9) \times 10^{-3}\approx \frac{1}{775}\,;
\end{equation}
namely, this difference is essentially driven by the mass ratio, being of the order of $(m_\pi/m_{\pi(1300)})^3 \approx 1/800$.

 The explanation, of course, stems from the significant similarity between the form factors of the pion and its excited state, as illustrated in Fig.\,\ref{fig:EFF} and Ref.\,\cite{Miramontes:2021exi}. This resemblance persists despite the substantial differences in their corresponding wave functions (see Fig. 1 and Ref.\,\cite{Li:2016dzv}). An analogous situation can be observed between the proton and the Roper resonance\,\cite{Segovia:2015hra}. In any case, from a basic intuitive standpoint, these outcomes can be understood as a result of the radial excitation having the same quantum numbers and flavor structure as the ground state.

\section{Towards the Kaon(1460) electromagnetic form factor and box contribution.}

\noindent
Returning to the estimations of the VMD models, in a manner similar to the pion, we can estimate the box contribution arising from the first radial excitation of the kaon, namely $K(1460)$, by employing a simple VMD formula:
\begin{widetext}
\begin{flalign}\label{eq:VMD2}
   F_{K(1460)}^{\text{VMD}}(Q^2) \hspace{-1mm} = \hspace{-1mm} 1- \hspace{-1mm} \frac{Q^2}{2} \hspace{-1mm} \Bigg[\frac{1}{m_{\rho}^2 + Q^2} + \frac{1}{3} \hspace{-1mm} \left(\frac{1}{m_{\omega}^2 +Q^2} \right) \hspace{-1mm} +\frac{2}{3} \hspace{-1mm} \left(\frac{1}{m_{\phi}^2 + Q^2}\right) \hspace{-1mm} 
   \Bigg]\,,
\end{flalign}
\end{widetext}
\hspace{-3mm}
with $m_{\rho},m_{\omega},m_{\phi}
$ being the vector meson masses
. We might include the contribution from the first excited multiplet of vector mesons, as in Eq.~(\ref{eq:VMD}). However, the complicated interplay between this and the contribution from the next heavier states seems rather non-trivial according to the wiggles observed in the BaBar study \cite{BaBar:2013jqz}, which motivates our simplified description in Eq.~(\ref{eq:VMD2}). Still, it satisfies the chiral limit expectations and falls off asymptotically as expected on QCD grounds. Employing this parameterization, we obtain
\begin{equation}
    a_{\mu}^{K(1460)-\text{box-VMD}} = -1.08  \times 10 ^{-13} .
\end{equation}
 While $a_\mu^{\pi-box}/a_\mu^{K-box}\sim30$ for the ground states, $a_\mu^{\pi_{1}-box}$ and $a_\mu^{K_{1}-box}$ have comparable order of magnitude. This is due to the large hierarchy between $m_\pi^2$ and $m_K^2$ and the consequent effect on the pseudo NG-boson propagators entering the evaluation of $a_{\mu}^{\textbf{P}-box}$, Eq.~(\ref{eq.master_formula}). On the contrary, $m_{\pi(1300)}^2\sim m_{K(1460)}^2$, yielding $a_\mu^{\pi(1300)-box}\sim a_\mu^{K(1460)-box}$, as the corresponding EFFs do not differ substantially in the region that dominates the integral.

As with the pion, a SDE/BSE-based calculation of the $K(1460)$ box contribution requires the computation of the corresponding BSA. Nevertheless, employing a RL or BRL truncation does not describe properly the mass for this radial excitation. In Ref.~\cite{Xu:2022kng}, it was argued that, in order to obtain a better description for the $K(1460)$ mass, the quark chromomagnetic moment plays an important role.  This would be the case for any system  in which the current masses of its valence quarks differ substantially. Such a calculation including the chromomagnetic moment together with the MCE will be carried out and presented elsewhere. 

\subsection{Modified interaction to fit Kaon(1460) mass}
Since we aim to compute the contribution of radially excited charged pion and kaon box diagrams to the anomalous magnetic moment of the muon, it is mandatory to have a satisfactory description of the first excited kaon state from the SDE/BSE formalism. As a first approach, we have employed an additional set of free parameters for the RL truncation, where $\eta = 1.55$ and $\Lambda = 0.92$. In this case, the excited kaon mass is fitted to,
\begin{eqnarray}
    m_{K(1460)} &=& 1.460~\text{GeV}\,.
\end{eqnarray}
Nevertheless, in this case the kaon ground state mass gets inflated, yielding $m_{K
^{\pm}} = 0.570$ GeV and $m_{K
^{\pm}} = 0.557$ GeV for RL and BRL truncations, respectively. For the time being, we might have to live with different sets of parameters to describe the ground and first radial excitation of the kaon.
We expect this drawback to be remedied on including the quark chromomagnetic term in the quark-gluon interaction vertex. For this exploratory calculation we compute the corresponding box contribution,

\begin{equation}
a_{\mu}^{K(1460)-\text{box}}(\text{RL}) = - 1.38  \times 10^{-13}.
\end{equation}
As discussed before, it is of the order of $a_\mu^{\pi(1300)-\text{box}}$.

\begin{figure*}[ht]
\centerline{%
\includegraphics[width=0.95\textwidth]{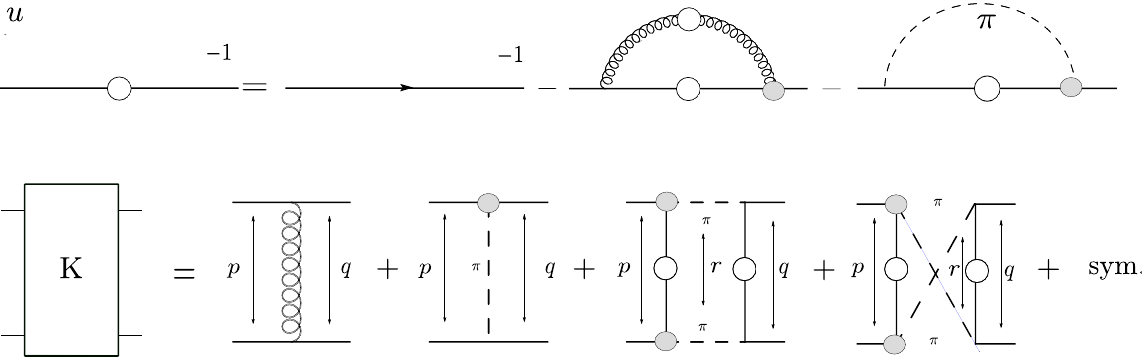}}
\caption{Truncations employed for the BSE interaction kernel $K$ (lower diagrams) and the quark SDE one (upper diagrams). 
In the lower panel, the diagrams on the right-hand side correspond to the RL, pion exchange, and the $s$-channel together with the $u$-channel pion decay contributions, respectively.}
\label{fig:kernels}     
\end{figure*}

\section{Conclusions and scope} \label{Conclusions}

We have presented the first calculation of a space-like EFF of the radial excitation of the pion by employing the combined framework of the SDE/BSE in a truncation that includes RL diagrams and MCE. This framework enables us to properly capture crucial aspects as regards the description of the $\pi(1300)$. Among others: the simultaneously inflated mass and the minuscule leptonic decay constant, the strict vanishing of the latter in the chiral limit, and the development of a node in the BSA. Furthermore, as opposed to the RL approximation, the BRL truncation permits an on par description of both the ground-state and first radial excitation within a unique set of parameters.

In addition, we have estimated the corresponding box-contribution to the muon's anomalous magnetic moment stemming from HLbL processes. We have compared our results with those obtained from a VMD model. As expected, our findings reveal that this contribution is much smaller when compared to the similar computation for the ground-state pion. The difference in the order of magnitude is driven, mostly, by the mass ratio $(m_\pi/m_{\pi(1300)})^3$. It is concluded that this occurs due to the substantial similarity between the EFF of the ground state and the radial excitation, which takes place despite the considerable differences at the BSA level, and which also translates into a similar spatial extension in both systems.

Extending our analysis on the $\pi(1300)$, we perform an exploratory study on its kaonic counterpart, $K(1460)$. A first approach shows that, despite the virtues of the BRL scheme, it is unable to describe the ground-state and the first radial excitation of the Kaon with a single set of parameters. The most probable reason lies in the difficulty of deriving truncations capable of properly capturing the flavor asymmetry. In this context, it is suggested to incorporate in the future the beyond RL pieces in Ref.\,\cite{Xu:2022kng}, which take into account the anomalous chromomagnetic moment of the quark. For these reasons, we estimate the box contribution arising from the $K(1460)$ by utilizing a VMD representation instead. It is worth noting that a comprehensive calculation within the SDE/BSE framework demands a more intricate truncation, a task we plan to undertake in future research.

\section{Acknowledgments}
A.~S.~Miramontes acknowledges {\em Consejo Nacional de Ciencia, Humanidades y Tecnología} (CONAHCyT), Mexico, for the financial support received through the program ``{\em Postdoctorados Nacionales por México}", additionally, has been partially funded by the “PROMETEO” programme of the “Generalitat Valenciana” grant CIPROM/2022/66, and by the Spanish “Agencia Estatal de Investigación”, MCIN/AEI/10.13039/501100011033, through the grants PID2020-113334GB-I00 and PID2023-151418NB-I00. A. Bashir thanks {\em Coordinación de la Investigación Científica}
of the {\em Universidad Michoacana de San Nicolás de Hidalgo} Grant No. 4.10., CONAHCyT grant CBF2023-2024-3544 and {\em Ayudas Beatriz Galindo}, Spain. P. Roig acknowledges Conahcyt (México) funding through project CBF2023-2024-3226 as well as Spanish support during his sabbatical year through projects MCIN/AEI/10.13039/501100011033, grant PID2020-114473GB-I00, and Generalitat Valenciana grant PROMETEO/2021/071. G.~Paredes-Torres acknowledges CONAHCyT, Mexico, for the financial support provided to him through the program ``{\em Beca de Posgrado en México}". The work of A. Bashir and K. Raya is supported by Spanish Ministry of Science and Innovation
(MICINN grant no. PID2022-140440NB-C22) and Junta
de Andaluc\'ia (grant P18-FR-5057).

\appendix
\section{Pion exchange interaction kernel}
The truncated Bethe-Salpeter interaction kernel which includes the exchange of explicit mesons as degrees of freedom, defined originally in~\cite{Fischer:2007ze,Fischer:2008sp}, is\,:
\begin{flalign}
& \hspace{-1mm} K^{(t)~ut}_{rs}(q,p;P) =\nonumber\\
&~~~~~~~\frac{C}{4} [\Gamma_{\textbf{P}}^j]_{ru} \left(\frac{p + q - P}{2}; p - q \right) [Z_2 \gamma^5]_{ts} D_{\textbf{P}}(p - q) \nonumber \\ \nonumber
 &~~~+\frac{C}{4} [\Gamma_{\textbf{P}}^j]_{ru} \left(\frac{p + q - P}{2}; q - p \right) [Z_2 \gamma^5]_{ts} D_{\textbf{P}}(p - q) \\ \nonumber
 &~~~+\frac{C}{4} [Z_2 \gamma^5]_{ru} [\Gamma_{\textbf{P}}^j]_{ts} \left(\frac{p + q + P}{2}; p - q \right) D_{\textbf{P}}(p - q) \\ 
 &~~~+\frac{C}{4} [Z_2 \gamma^5]_{ru} [\Gamma_{\textbf{P}}^j]_{ts} \left(\frac{p + q + P}{2}; q - p \right) D_{\textbf{P}}(p - q)~,\label{eq:BSEkernel_tchannel}
\end{flalign}

\noindent together with the corresponding truncation of the quark SDE
\begin{eqnarray}
&& \hspace{-4mm} S^{-1}(p) = S^{-1}(p)^{RL} - \frac{3}{2} \int_q \Bigg[Z_2 \gamma_5 S(q) \Gamma_{\textbf{P}}\left(\frac{p+q}{2}, q-p\right) \nonumber \\
&& \hspace{+8mm} + Z_2 \gamma_5S(q)\Gamma_{\textbf{P}}\left(\frac{p+q}{2}, p-q\right)\Bigg] \frac{D_{\pi}(k)}{2}~,\label{eq:quarkDSE_tchannel}
\end{eqnarray}
where \textbf{P}
$=\pi$ is the meson under study and $S^{-1}(p)^{RL}$ is the RL truncation with the gluon-mediated interaction term. 
In Eqs.~\eqref{eq:BSEkernel_tchannel} and \eqref{eq:quarkDSE_tchannel} the pion propagator is taken as 
$D_{\pi}(k) = {(k^2 + m_{\pi}^2)^{-1}}$. 

Additionally, $C$ in Eq.~\eqref{eq:BSEkernel_tchannel} is a flavor factor, as discussed in detail in~\cite{Miramontes:2021exi,Miramontes:2019mco}; when we use $C=3/2$ to compute the quark-photon vertex, it leads to a small violation of the AxWGTI. For all the calculations in this paper we have used instead $C=-3/2$ which satisfies the AxWGTI but leads to a small violation of the VGWTI (less that $1\%$). Herein, the quark-meson vertex $\Gamma_\pi$ is taken to be the full pion BSA. On the other hand, the exchange of the pions in the interaction kernel will also appear in the $s$- and the $u$-channels~\cite{Fischer:2007ze}. They read as follows, 

\vspace{20mm}
\begin{widetext}
\flaligne{
K^{(s)~he}_{da}(q,p,r;P)=~& \frac{C}{2}~D_{\textbf{P}}\left(\frac{P + r}{2}\right) D_{\textbf{P}}\left(\frac{P - r}{2}\right)~\left[[Z_2\gamma_5]_{dc} S_{cb}\left(p - \frac{r}{2}\right)[Z_2\gamma_5]_{ba}\right.\nonumber\\
&~~~~\times[\Gamma_{\textbf{P}}^j]_{hg} \left(q - \frac{P}{4} - \frac{r}{4}; \frac{r - P}{2} \right) S_{gf}\left(q - \frac{r}{2}\right)[\Gamma_{\textbf{P}}^j]_{fe} \left(q + \frac{P}{4} - \frac{r}{4}; -\frac{P + r}{2} \right)\nonumber\\
~&+~[\Gamma_{\textbf{P}}^j]_{dc} \left(p + \frac{P}{4} - \frac{r}{4}; \frac{P + r}{2} \right) S_{cb}\left(p - \frac{r}{2}\right)[\Gamma_{\textbf{P}}^j]_{ba} \left(p - \frac{P}{4} - \frac{r}{4}; \frac{P - r}{2} \right)\nonumber\\
&~~~~\left.\times[Z_2\gamma_5]_{hg} S_{gf}\left(q - \frac{r}{2}\right)[Z_2\gamma_5]_{fe}\right] \,, 
\label{eq:BSEkernel_schannel}}
\end{widetext}
 where $r$ is an additional integration momentum in the BSE. A similar term can be written for the $u-$ channel diagram. The resulting truncation of the BSE kernel as well as the quark SDE are depicted in Fig.~\ref{fig:kernels}.

 As discussed in Ref.~\cite{Miramontes:2019mco}, the inclusion of the interaction kernel in Eq.~\eqref{eq:BSEkernel_tchannel} does not satisfy the AxWGTI and the VWGTI at the same time. To effectively capture the NG mode characteristics of the pion, we have chosen $C=-3/2$ in the  kernels described above. This allows the AxWGTI to be faithfully upheld. In relation to the VWGTI, a $1\%$ deviation in the proper normalization of the EFF, $F_{\textbf{P}}(Q^2=0) = 1$, points to a marginal violation of this identity. This also highlights the necessity of supplementing the IA to include additional diagrams. Therefore, consider the electromagnetic in Eq.\,\eqref{eq:Current}, expressed as follows:
\begin{equation}
    \label{eq:symprescur}
    J^{\mu} = \bar{\Psi}_{\textbf{P}}^f G_0 (\mathbf{\Gamma}^{\mu}_{\text{BRL}} +\mathcal{R}^{\mu}) G_0 \Psi_{\textbf{P}}^i ~;
\end{equation}
as before, $\mathbf{\Gamma}^{\mu}$ denotes an IA computation, while the subscript $\text{`BRL'}$ means that all required components have been derived within the BRL truncation. For its part, $\mathcal{R}^{\mu}$ corresponds to the symmetry-restoring term:
\begin{equation}
    \label{eq:symres}
    \mathcal{R}^{\mu}:= [1-\mathbf{\Gamma}^{\mu}_{\text{BRL}}]_{Q^2=0}\times \mathbf{\Gamma}^{\mu}_{\text{RL}}\,.
\end{equation}
This piece produces imperceptible variations to the original result, on the order of $\lesssim 1 \,\%$. Moreover, since the RL approximation ensures $F_{\textbf{P}}(Q^2=0)=1$, the construction defined in Eq.\,\eqref{eq:symprescur} and Eq.\,\eqref{eq:symres} guarantees that this holds true for the current BRL scheme as well.


\section{Analytic continuation to complex plane}
\label{Cauchy}
In order to compute the meson BSA we require the knowledge of the quark propagator for complex momentum. In Euclidean space-time, the total momentum $P$ is parametrized as $P = (0,0,0,i M )$, with $P^2=-M^2$. In this case, the quark propagator $S(q \pm P/2)$ is sampled within complex parabolas defined by
\begin{equation}
    q_{\pm} = q^2 - M^2 +i z \sqrt{q^2} \sqrt{M^2} \,,
    \label{eq:parabola}
\end{equation}
where $q$ is the relative quark momentum, $M$ is the bound state mass and $z$ is the angle between $q$ and $P$. The Eq.~\eqref{eq:parabola} forms a parabola in the complex plane centered at $M^2/4$.
To solve the quark SDE in the complex plane one commonly employed technique is the Cauchy interpolation method. The Cauchy integral formula for a closed contour $\gamma$ and a complex number $z_0$ reads as
\begin{equation}
    f(z_0) = \frac{1}{2 \pi i} \oint_\gamma \frac{f(z)}{z -z_0}dz
    \,.
    \label{eq:cauchy}
\end{equation}
We can rewrite Eq.~\eqref{eq:cauchy} as
\begin{equation}
      f(z_0) = \frac{\oint_\gamma \frac{f(z)}{z -z_0}dz}{\oint_\gamma \frac{1}{z-z_0}} \,.
      \label{eq.cauchy2}
\end{equation}
Employing Eq.~\eqref{eq.cauchy2}, we can numerically calculate the dressing function of the quark propagator in Eq.~\eqref{QuarkProp} by solving its SDE. Nevertheless, the quark propagator exhibits a pair of complex conjugate poles in the complex plane. It limits the region where the Cauchy interpolation can be used. Beyond the position of the singularities of the quark propagator we employ a parameterization consisting of the following complex conjugate pair {\em Ansatz}, which is a common practice\,: 
\begin{eqnarray}
S(p) &=& -i \slashed{p} \sigma_v(p^2) + \sigma_s(p^2) \, ,\nonumber \\ 
\sigma_v (p^2) &=& \sum_{i}^n \left[\frac{\alpha_i}{p^2 + m_i} + \frac{\alpha_i^\ast}{p^2 + m_i^\ast}\right]  \, , \nonumber \\
\nonumber \\
\sigma_s (p^2) &=& \sum_{i}^n \left[\frac{\beta_i}{p^2 + m_i} + \frac{\beta_i^\ast}{p^2 + m_i^\ast}\right] \,,
\end{eqnarray}
where the parameters $m_i$, $\alpha_i$, $\beta_i$ can be obtained by fitting the corresponding quark SDE solution 
along the $p^2$ real axis or, alternatively, on a  parabola in the complex plane that does not enclose the poles. We use two pairs of complex conjugate poles as these are enough to provide a sufficiently precise fit for the quark propagator.

\section{Chebyshev expansion}\label{Cheby}
The process of calculating the dressing functions from the homogeneous BSE can be simplified by factorizing the angular dependence and subsequently expanding it in terms of Chebyshev polynomials of the second kind. This method allows for a more efficient representation and computation of the dressing functions. For instance, consider the following expression:
\begin{equation}
c_i(q^2,p^2) = \sum_j c_{ij}(q^2,p^2) P_j (z) \,.
\end{equation}
Here, $P_j(z)$ denotes the Chebyshev polynomials. These are a sequence of orthogonal polynomials, which are used extensively in numerical analysis due to their convenient properties, providing an optimal choice for interpolation and approximation of functions over a given interval.

The variable $z$ in the equation is the cosine of the angle, calculated from the dot product of $q$ and $p$, where $q$ represents the quark momentum and $p$ symbolizes the total momentum. Importantly, $z$ falls within the range $(-1,1)$, which is the standard interval for the Chebyshev polynomials. By employing this factorization and expansion approach, we can simplify the calculation of the dressing functions, making it more efficient and manageable, especially for the extensive computational tasks at hand.

\bibliographystyle{unsrt}
\bibliography{main}

\end{document}